Letter

# Projections of H→ $\tau\tau$ cross-section at FCC-ee

**S. Giappichini**[1,a], **M. Klute**[1], **M. Presilla**[1], **X. Zuo**[1,2], **M. Cepeda**[3]

[1] Institute for Experimental Particle Physics (ETP), Karlsruhe Institute of Technology (KIT), Karlsruhe, Germany
[2] École Polytechnique Fédérale de Lausanne (EPFL), Lausanne, Switzerland
[3] Centro de Investigaciones Energéticas, Medioambientales y Tecnológicas (CIEMAT), Madrid, Spain

Received: 8 June 2026 / Accepted: 4 August 2026

**Abstract** The Future Circular Collider (FCC) stands at the forefront of the European Strategy for Particle Physics as the future flagship project at CERN. The H→ $\tau\tau$ decay, featuring a large branching ratio, clean identification in the FCC-ee environment, and the possibility to reconstruct tau spin-related observables, is an excellent channel to measure Higgs boson properties. This work shows the expected precision for the H→ $\tau\tau$ cross-section measurement at the FCC-ee in the ZH production mechanism at $\sqrt{s}$ = 240 GeV and $\sqrt{s}$ = 365 GeV, as well as via the vector boson fusion process at $\sqrt{s}$ = 365 GeV. Furthermore, we explore and evaluate a set of methods for reconstructing tau decays. These techniques are critical for unlocking the full physics potential of the FCC-ee and for improving the understanding of tau-related observables in both Standard Model measurements and New Physics searches. The obtained results significantly enhance the FCC-ee outlook in the H→ $\tau\tau$ channel, improving it by at least an order of magnitude compared to the current uncertainties on the cross-sections measured at the LHC.

## 1 Introduction

After more than a decade of measurements at the LHC, Higgs boson couplings to Standard Model (SM) particles are now probed with percent-level precision across multiple production and decay channels, providing a unique window on the dynamics of the Electroweak Symmetry Breaking mechanism and possible new physics beyond the SM. In this context, precision studies of Higgs boson Yukawa interactions with charged leptons are especially powerful, since small deviations from the SM pattern can signal new degrees of freedom at higher scales. Among these, the H→ $\tau^+\tau^-$ decay plays a central role: it is the most accessible leptonic Higgs boson decay mode at hadron colliders, with a relatively large branching fraction and a more favorable background environment compared to H→ $b\bar{b}$. The ATLAS and CMS Run I combination, obtained from proton–proton collisions at $\sqrt{s} = 7$ and 8 GeV, marked the first observation of the H→ $\tau^+\tau^-$ decay [1], while the first observation from a single experiment was obtained by the CMS experiment using data collected at a center-of-mass energy of 13 TeV [2].

The interest in the exploration of the Higgs boson sector with new dedicated colliders, alongside top and electroweak physics, has been recently formalized by the European Strategy for Particle Physics Update [3–5]. These machines offer a uniquely clean electron–positron collision environment, combining unprecedented luminosity with excellent signal-to-background ratios, and thereby enable major advances in precision measurements on comparatively short data-taking timescales.

The $e^+e^-$ Future Circular Collider (FCC-ee) [6,7] is one of such proposed machines to be built at CERN in the coming years. The core of the FCC-ee program is to measure Higgs boson production inclusively [8,9] with a dedicated run at $\sqrt{s}$ = 240 GeV, where about 2 million Higgs bosons would be produced in association with Z bosons (ZH) for an integrated luminosity of 10.8 ab$^{-1}$. Additionally, a run around the $t\bar{t}$ threshold, at $\sqrt{s}$ = 365 GeV with an expected collected luminosity of 3 ab$^{-1}$, would also make accessible Higgs boson production from vector boson fusion (VBF). It would then be the first time that the Higgs boson is produced from $e^+e^-$ collisions, allowing for model-independent measurements of its properties [10].

The main goal of this letter is to estimate the achievable relative uncertainty on the cross-section of the Higgs boson decay into a pair of taus in the ZH production channel at FCC-ee, at $\sqrt{s}$ = 240 GeV and $\sqrt{s}$ = 365 GeV, as well as from VBF at the top threshold, with the respective Feynman diagrams shown in Fig. 1. The current measurements

[a] e-mail: sofia.giappichini@cern.ch (corresponding author)

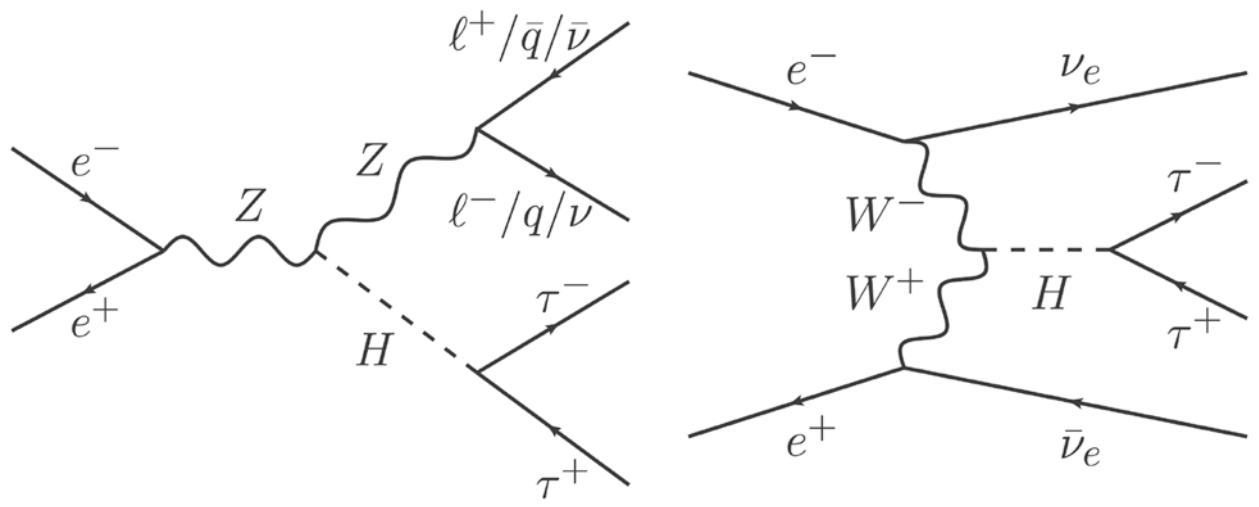

**Fig. 1** On the left: Feynman diagram of ZH events in $e^+e^-$ collisions, where the Z boson decays to leptons (electrons or muons considered in this study), quark or neutrino pairs, and the Higgs boson decays to a tau pair. On the right: Feynman diagram of VBF events (mediated by W bosons) in $e^+e^-$ collisions, where the Higgs boson decays to a tau pair

by the LHC experiments in these channels on the product $\sigma \cdot BR(H \rightarrow \tau^+\tau^-)$, normalized to the SM predictions, are respectively $\mu_{ZH} = 1.89^{+0.65}_{-0.56}$ and $\mu_{VBF} = 0.86^{+0.17}_{-0.16}$ from CMS [11], and $\mu_{ZH} = 1.00^{+0.62}_{-0.59}$ and $\mu_{VBF} = 1.00^{+0.21}_{-0.18}$ from ATLAS [12], where the signal strength $\mu$ is defined as the ratio of the signal yield observed to that predicted by the SM. This represents relative uncertainties of the order of 60% in the first channel and 20% in the second. The projections for HL-LHC, the next upgrade of the LHC with $\mathcal{L}_{int} = 3$ ab$^{-1}$ and $\sqrt{s} = 14$ GeV, lower the value to 14% in the VH channel and 4-8% in VBF [13,14]. A different way to interpret the measurement is to consider the Higgs boson coupling modifiers in the $\kappa$ framework [15,16], where $\kappa_i$ parametrizes the measurable ratio of the Higgs boson coupling strength to particle $i$ relative to the SM prediction. In this scheme, the current value for $\kappa_\tau$, which is related to the H$\rightarrow \tau^+\tau^-$ vertex, is $0.92 \pm 0.08$, giving a relative precision of 8.6% from CMS [11], and $0.93 \pm 0.07$, corresponding to 7.5% relative precision, from ATLAS [12]. The corresponding estimation at HL-LHC improves the relative precision to 1.9% [17].

In the scope of this study, we also review and compare methods of tau reconstruction at future $e^+e^-$ colliders, necessary to achieve a high precision in the analyzed channel. Among the charged leptons, the tau remains the least precisely characterized, largely because of its numerous, predominantly hadronic decay modes, with the Belle II collaboration recently reporting the most precise measurement of its mass [18]. It is expected that more than $10^{11}$ tau pairs will be produced at FCC-ee from the Z-pole stage alone ($\sqrt{s} \simeq 91$ GeV with $\mathcal{L}_{int} = 205$ ab$^{-1}$ [7]), rivaling both dedicated flavor factories, on account of the high luminosity achievable, and hadron colliders, where information loss is caused by the presence of underlying event, pileup, and unknown partonic center-of-mass energy. This would result in the improvement by several orders of magnitude of the current limits on tau observables [19,20].

## 2 Event simulation and reconstruction

We use Monte Carlo (MC) samples for the signal and the SM backgrounds, simulated at Leading Order with different MC generators, including a parameterization of initial and final state radiation. The SM signals and all the relevant Higgs boson decays from ZH production (ZZ, $W^+W^-$, gg, $b\bar{b}$, $c\bar{c}$, and $s\bar{s}$), are generated with Whizard v. 3.0.3 [21,22]. The same generator is used to simulate single Z boson production to leptons, $\gamma\gamma \rightarrow \ell^+\ell^-$, $e\gamma \rightarrow eZ$ with Z decaying into electrons or muons, and $ee \rightarrow \nu\bar{\nu}Z$. The hadronization of partons is simulated with Pythia6 [23]. The single Z boson decay to quarks, $e^+e^- \rightarrow ZZ$, and $e^+e^- \rightarrow W^+W^-$, are instead produced with Pythia8 [24]. For the analysis performed at $\sqrt{s} = 365$ GeV, the additional background coming from $t\bar{t}$ is considered and simulated with Pythia8.

A simulation of the IDEA detector concept [25], one of the main designs proposed for FCC-ee based on a short-drift wire chamber and a dual-readout calorimeter, is performed. It relies on the fast-simulation package DELPHES v.3.5.1pre05 [26]. Here, the reconstruction efficiencies are defined so that all leptons have at least energy of 2 GeV, $p_T > 0.1$ GeV, and $|\eta| < 2.56$, photons have the same energy requirement and $|\eta| < 3.0$, and charged hadrons pass $p_T > 0.1$ GeV. The FastJet package [27] is used to cluster particles into jets with the $e^+e^-$ exclusive Durham algorithm, fixing $n_{jets}$ to the expected jet multiplicity (including hadronic tau decays) of the signal topology under study. The exclusive clustering was proven to be the preferred approach in lepton collider analyses, as the absence of underlying event and pileup allows the jet count to be fixed directly to the known hard-parton multiplicity of the signal process.

The analysis is organized into nine categories based on the Z boson and tau decay modes. The defining characteristic of each category is the number of leptons and hadronically decaying particles present in the final state. Categories with no hadronic decays bypass jet clustering entirely. For the rest, leptons are excluded from the jet constituents, and $n_{jets}$ is set equal to the number of expected hadronic decays. In the specific case of $Z \rightarrow q\bar{q}$, we only consider leptons with $p > 20$ GeV and isolation $p_{tot}/p_{lepton} < 0.25$, where $p_{tot}$ is the sum of all particle's momenta in a cone of $0.01 < \Delta R < 0.5$ around the lepton, since low-momentum leptons could arise from the quark decays. The selection requirements and clustering strategy for each category are summarized in Table 1. The VBF channel follows the Z$\rightarrow \nu\bar{\nu}$ selection.

Leptonic Z and tau decays ($\tau_\ell$) are both identified directly from the isolated electrons and muons. To distinguish them in case of ambiguity, we first identify the opposite-charge, same-flavor lepton pair whose invariant mass is the closest to the nominal Z boson mass and assign it to the Z candidate. The remaining leptons are then associated with the tau candidates.

**Table 1** Definition of final object selection and clustering in each category of the analysis. Leptons, $\ell$, are either electrons or muons, $q$ stands for all quarks except the top quark, and $\nu$ includes all neutrino flavors

| | $H \to \tau_h\tau_h$ | $H \to \tau_\ell\tau_h$ | $H \to \tau_\ell\tau_\ell$ |
|---|---|---|---|
| $Z \to \ell^+\ell^-$ | Two leptons (opposite charge same flavor), two jets | Three leptons (one opposite charged pair), one jet | Four leptons (opposite charged pairs with one same flavor pair), no clustering |
| $Z \to \nu\bar{\nu}$ | No leptons, two jets | One lepton, one jet | Two leptons (opposite charge), no clustering |
| $Z \to q\bar{q}$ | Four jets | One isolated lepton, three jets | Two isolated leptons, two jets |

Hadronic tau decays ($\tau_h$) are constructed starting from jets with the procedure described in the following section, while also separating them from quark jets.

Missing energy, $E_{miss}$, is identified by subtracting the momenta of all visible particles from the center-of-mass energy of the collision, which, in the conditions of a lepton collider, is expected to be precisely known and thus reliable.

## 3 Tau reconstruction at the FCC-ee

This section compares two techniques for reconstructing hadronic tau decays at FCC-ee. In addition to their use in the analyses presented in this letter, these methods are broadly applicable to similar processes, such as $Z \to \tau^+\tau^-$ events, and are crucial to maximizing the physics reach of future collider experiments. We examine a machine-learning-based approach and a method that explicitly assigns events to individual tau decay modes.

*ParticleNet reconstruction.* The ParticleNet algorithm [28], based on a graph neural network architecture for jet flavor identification, was previously trained in the collision environment of the IDEA detector on di-jet ZH events at $\sqrt{s} = 240$ GeV, with the Z boson decaying into a pair of neutrinos [29], and offers the capability of distinguishing taus from jets by providing, for each jet, a classification score corresponding to the probability that it originates from a given quark flavor, a gluon, or a tau. In this case, a jet is identified as a tau if its tau score exceeds 0.5, a threshold chosen to reject most quark and gluon jets while retaining a high tau identification efficiency: the fake rate is $3.97 \pm 0.13\%$ in the $Z \to q\bar{q}$ sample and the identification efficiency is $99.954 \pm 0.002\%$ in $\nu\bar{\nu}H(\to \tau_h\tau_h)$ events. This value could be tuned in the future to achieve better purity at the expense of efficiency. The algorithm does not perform a tau decay mode identification on its own, but the jet constituents information is still available for further applications.

*Explicit reconstruction.* To identify the exact tau decay mode, a function is implemented to scan through the single jet constituents and identify the corresponding tau products. First of all, jets containing electrons or muons are immediately tagged as quark jets. For the remaining, the four-vectors are based on the constituents spatially close ($\Delta\theta < 0.2$) to the leading charged particle of the jet. Further requirements check for the sum of the charges to be $\pm 1$ and mass smaller than 3 GeV, to be consistent with the hypothesis of a tau origin. The number of charged and neutral constituents, either neutral kaons or photons, is used to infer the tau decay mode, assigning a separate value to the rejected jets. The reconstruction of neutral pions from photons is not implemented in this scope.

Table 2 shows the efficiency of both reconstruction methods, matching the reconstructed taus with the true ones in a cone of $\Delta R = 0.2$. It is expected that the efficiency drops significantly when the Z decays hadronically, as these events are less clean. The efficiencies of the two methods are comparable, so the choice is mainly motivated by the need to identify the tau decay modes. The two methods could be combined, performing decay-mode identification on tau-tagged jets. This has not been explored in this analysis, as the natural combination of the methods would result in a loss of efficiency, with a potential gain in purity that is not relevant for the scope of this study.

In Fig. 2, the confusion matrix for the individuated tau decay mode against the generated ones is shown for the cases where the truth-matching is successful. This illustrates an almost perfect efficiency for all major decay modes. The explicit reconstruction of neutral pions is not performed due to the challenge of separating and reconstructing two or more photons in their decays.

In electron–positron collisions, it is possible to recover information about neutrinos in tau pairs with up to twofold ambiguity by considering the event's kinematics. However, some considerations need to be made. Firstly, either the center-of-mass of the di-tau system or the recoil system needs to be well known, meaning the missing energy in the event is exclusively related to the tau neutrinos. Secondly, the taus need to have enough separation to be reconstructed in two distinct jets. Finally, the tau mass is set a priori, and neutrinos are treated as massless particles. Examples can be found in Refs. [30,31]. In this analysis, we do not exploit the full tau momentum information, as it does not significantly improve tau identification. Instead, it is primarily relevant for extracting sensitivity to new physics in the Higgs boson sector and for studying the tau Yukawa coupling; therefore, these methods will not be discussed further.

**Table 2** Efficiency of hadronic tau decay reconstruction with the ParticleNet algorithm and explicit reconstruction at $\sqrt{s} = 240$ GeV and 365 GeV. The values correspond to matching the reconstructed tau to the generated one in a cone of $\Delta R < 0.2$. The efficiencies for different quark flavors have been combined since they are similar to each other

| Efficiency [%] | ParticleNet | | Explicit Reconstruction | |
|---|---|---|---|---|
| | 240 GeV | 365 GeV | 240 GeV | 365 GeV |
| $Z\to \nu\bar{\nu},\ H \to \tau^+\tau^-$ | $95.03 \pm 0.03$ | $93.67 \pm 0.03$ | $89.02 \pm 0.04$ | $88.80 \pm 0.04$ |
| $Z\to e^+e^-,\ H \to \tau^+\tau^-$ | $86.69 \pm 0.08$ | $81.43 \pm 0.06$ | $86.35 \pm 0.08$ | $84.14 \pm 0.06$ |
| $Z\to \mu^+\mu^-,\ H \to \tau^+\tau^-$ | $90.69 \pm 0.07$ | $87.17 \pm 0.05$ | $87.37 \pm 0.08$ | $85.95 \pm 0.06$ |
| $Z\to q\bar{q},\ H \to \tau^+\tau^-$ | $61.90 \pm 0.07$ | $74.20 \pm 0.03$ | $78.13 \pm 0.06$ | $80.92 \pm 0.03$ |

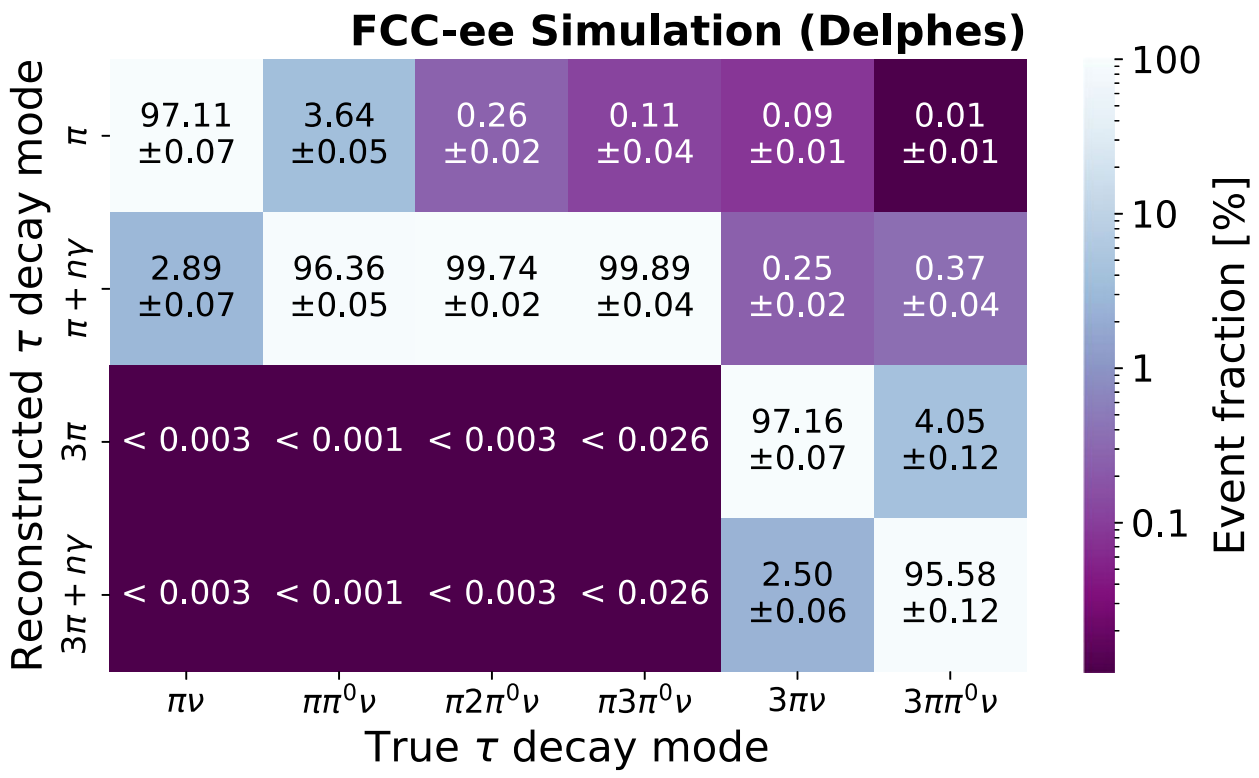

**Fig. 2** Confusion matrix of identified tau decay modes against truth level ones from the explicit reconstruction method for $Z\to \nu\bar{\nu}$, $H\to \tau^+\tau^-$ events at $\sqrt{s} = 240$ GeV, where two jets are clustered in the events. Only taus that are matched in a cone of $\Delta R = 0.2$ to the true ones are considered. Upper bounds are calculated with the Clopper–Pearson method

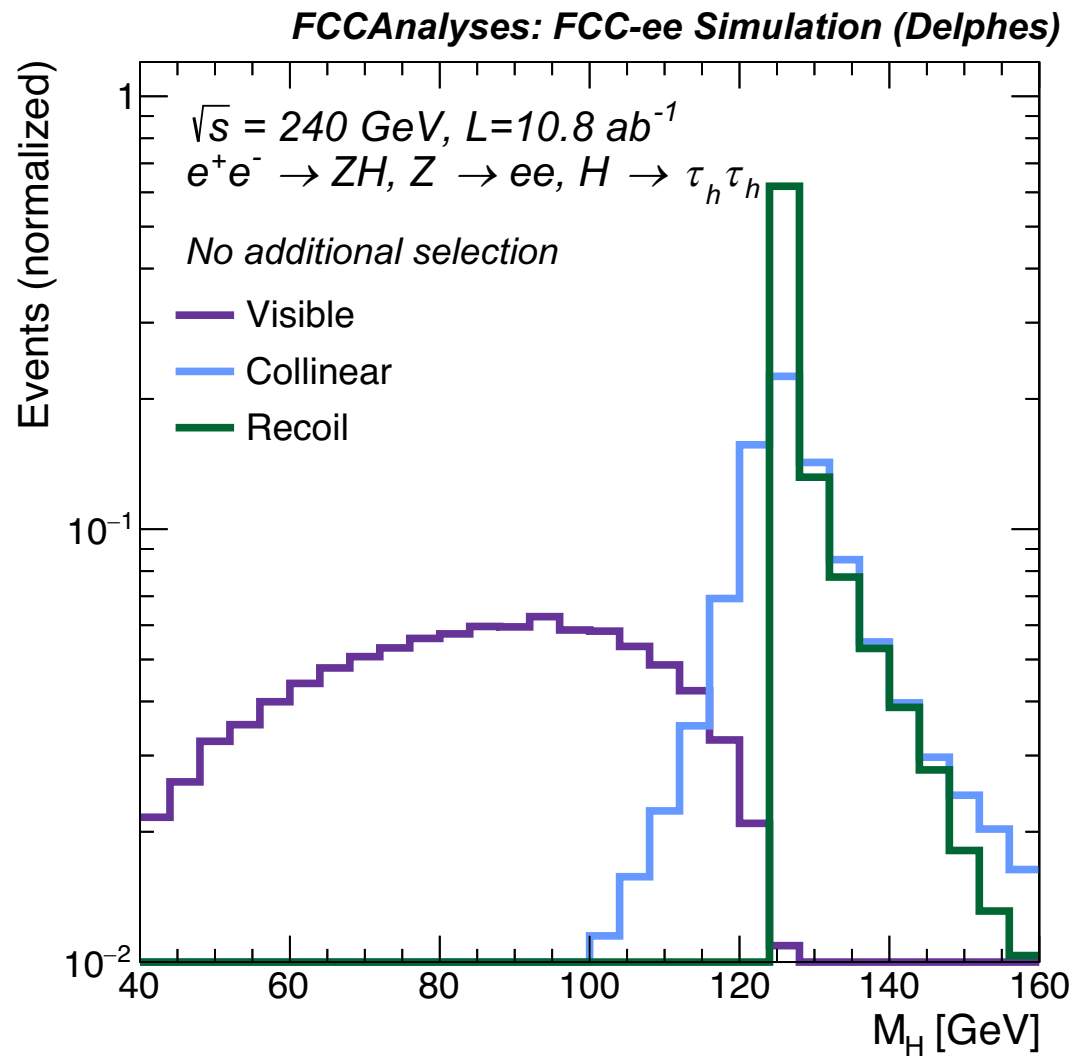

**Fig. 3** Normalized distributions of the reconstructed Higgs boson mass for $Z \to e^+e^-$, $H \to \tau^+\tau^-$ events for the visible, collinear approximation, and recoil mass techniques. ParticleNet is employed on two exclusive jets reconstructed in the event to identify taus

Since we are interested in studying the Higgs boson, we also compare different approaches to obtain its mass. First of all, the invariant mass of the tau pair is calculated using the collinear approximation [32], $M_{collinear}$. Then, the mass of the system recoiling against the tau lepton pair is extracted from the momenta of the decay products of the Z and the total energy-momentum conservation. Finally, we compare them to the invariant mass of the visible decay products of the taus. From Fig. 3, the visible mass of the two taus carries the least amount of information, given the missing energy of the neutrinos. The collinear and recoil masses are capable of identifying the Higgs boson precisely, with the latter having an advantage shown by the very narrow peak with respectively $60.97 \pm 0.12\%$ and $83.74 \pm 0.09\%$ of events in a region between 120 and 132 GeV.

**Table 3** The second, third, and fifth columns show the relative uncertainty on $\sigma_{ZH} \times \mathcal{B}(H \to \tau^+\tau^-)$ (ZH), while the fourth refers to $\sigma_{\nu_e\bar{\nu}_e H} \times \mathcal{B}(H \to \tau^+\tau^-)$ (VBF). The center-of-mass energy is $\sqrt{s} =$ 240 GeV in the second column, $\sqrt{s}$ =365 GeV in the third and fourth, while the last column presents the combination of the two. All columns indicate the corresponding integrated luminosity assumed in the analysis. These values are extracted by considering the ParticleNet tau reconstruction with exclusive jet clustering and a BDT event discrimination

| $\sqrt{s}$ <br> $\mathcal{L}$ | 240 GeV <br> 10.8 ab$^{-1}$ | 365 GeV <br> 3 ab$^{-1}$ | | 240+365 GeV <br> 13.8 ab$^{-1}$ |
|---|---|---|---|---|
| | ZH | ZH | VBF | ZH |
| $H\to \tau^+\tau^-$ | $\pm$ 0.57% | $\pm$ 1.17% | $\pm$ 8.48% | $\pm$ 0.51% |

## 4 Cross-section at $\sqrt{s}$=240 and 365 GeV

From the discussion in Sect. 3, we choose to proceed in this analysis with the ParticleNet tau reconstruction, considering that we do not exploit the decay mode information. Primary results obtained with the explicit tau reconstruction showed similar conclusions.

*BDT classifier.* A boosted decision tree (BDT) classifier is trained in each signal category and at the different energies to focus on the specific kinematics, to enhance the signal-to-background separation compared to a traditional cut-based

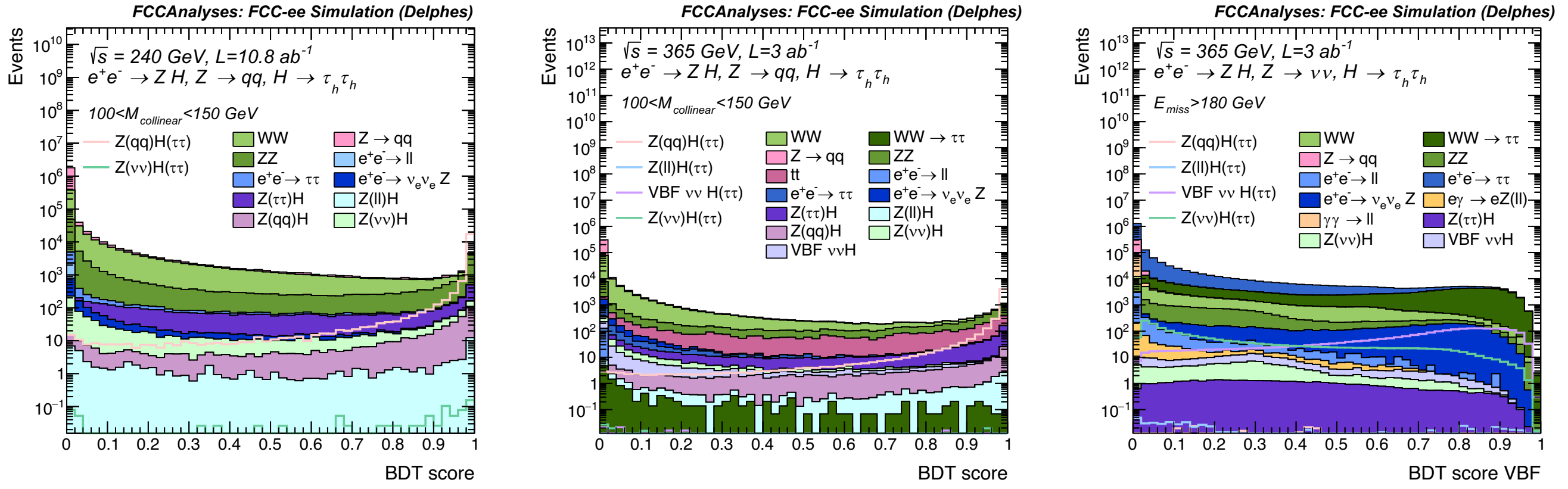

**Fig. 4** On the left: BDT score distribution for Z→ $q\bar{q}$, $H \rightarrow \tau_h^+\tau_h^-$ at $\sqrt{s} = 240$ GeV. On the center: BDT score distribution for Z→ $q\bar{q}$, $H \rightarrow \tau_h^+\tau_h^-$ at $\sqrt{s} = 365$ GeV. On the right: BDT score distribution for VBF $H \rightarrow \tau_h^+\tau_h^-$ at $\sqrt{s} = 365$ GeV

event selection. The BDT features include the di-tau $\Delta\phi$, $\Delta\eta$, $\Delta R$, missing four-momentum, reconstructed Z and Higgs boson four-momentum. The $M_{recoil}$ and $M_{collinear}$ features are also included as the most relevant discriminators, except for the $Z \rightarrow \nu\bar{\nu}$ and VBF categories where these can not be evaluated. Each BDT has a size of 200 trees and a depth of two. Given the limited event statistics, the full samples are used both for training and for the final analysis, after verifying the absence of over-training by comparing the BDT output distributions on training and testing subsets. For Z→ $q\bar{q}$ and Z→ $\ell^+\ell^-$, the BDTs are trained and then applied to events with $100 < M_{collinear} < 150$ GeV for both energies, while for Z→ $\nu\bar{\nu}$ the events have $E_{miss} > 100$ GeV. At the higher center-of-mass energy, a sizable contribution in the $\nu\bar{\nu}\tau^+\tau^-$ final state comes from the VBF channel, so a multi-class BDT is used to separate the two production modes, with the major differences being found in the missing transverse momentum and missing energy due to the differences between ZH and VBF topologies. In this case, the event selection for the training is adjusted to $E_{miss} > 180$ GeV to reflect the different kinematics. Figure 4 shows the BDT score distributions of the most sensitive channels in each separate energy and process.

*Results.* The cross-section measurement is extracted with the CMS Combine tool [33]. The maximum likelihood fit is performed using the BDT signal and background distributions, with four times as many bins as shown in Fig. 4. We consider a log-normal uncertainty of 1% on the cross-section of the background processes $e^+e^- \rightarrow ZZ$, $e^+e^- \rightarrow W^+W^-$, Drell–Yan, ZH, and photon-induced processes. Statistical uncertainty from the limited number of simulated events is not considered, as it does not significantly affect the result. Table 3 shows the results for the combination of each final state in the respective energies.

To extract the relative precision on $\kappa_\tau$, we consider the relations [15,16]:

$$\sigma_{ZH} \times \mathcal{B}(H \rightarrow \tau^+\tau^-) \propto \frac{\kappa_\tau^2\kappa_Z^2}{\Gamma_H},$$

$$\sigma_{\nu_e\bar{\nu}_e H} \times \mathcal{B}(H \rightarrow \tau^+\tau^-) \propto \frac{\kappa_\tau^2\kappa_W^2}{\Gamma_H},$$

where $\Gamma_H$ is the width of the Higgs boson and $\kappa_Z$ and $\kappa_W$ are the coupling modifiers of the ZZH and WWH vertices. At the FCC-ee, the relative precision on these values is expected to be around $\delta\kappa_Z = 0.1\%$, $\delta\kappa_W = 0.29\%$, and $\delta\Gamma_H = 0.78\%$ [7]. Given this, we expect the relative precision on $\delta\kappa_\tau = 0.47\%$ from the ZH production mode, while the VBF channel gives a worse result ($\delta\kappa_\tau = 4.26\%$) since it's limited by the higher background contribution.

## 5 Conclusions

In this letter, we have presented a first analysis of H→ $\tau^+\tau^-$ cross-section uncertainty in ZH events at $\sqrt{s} = 240$ GeV and ZH and VBF events at $\sqrt{s} = 365$ GeV for the $e^+e^-$ Future Circular Collider (FCC-ee). In particular, the relative uncertainty expected at FCC-ee for $\sigma_{ZH} \times \mathcal{B}(H \rightarrow \tau^+\tau^-)$ and $\sigma_{\nu_e\bar{\nu}_e H} \times \mathcal{B}(H \rightarrow \tau^+\tau^-)$ is obtained with a fast simulation of the IDEA detector concept, including all relevant background events and Z decay modes. We have also developed an algorithm to reconstruct hadronic tau decays that has been compared to a ParticleNet algorithm trained to recognize the jet flavor. Finally, a BDT classifier is employed to get higher discriminating power between signal and background.

The results of this study demonstrate that the FCC-ee could reach sub-percent precision of H→ $\tau^+\tau^-$ cross-section in the ZH production, significantly improving the current measurement by orders of magnitude, also considering the interpretation in the $\kappa$ framework. In contrast, in the VBF channel, the gain with respect to the LHC is less pronounced due to the lower expected number of events. Nonetheless, this illustrates how a future specialized Higgs boson factory can provide valuable insights into the Higgs boson sector.

**Acknowledgements** The work of S.G., M.K., and M.P. is supported by the Alexander von Humboldt-Stiftung. The work of M.C. is supported by grants PID2021-122134NB-C21 and CNS2023-144781 funded by MICIU/AEI /10.13039/501100011033 and FEDER, UE and European Union NextGenerationEU/PRTR.

**Data Availability Statement** This data will be made available on reasonable request. [Authors' comment: The analysis is based on large simulated dataset which can not be easily made available, with the relevant information given in the letter. The data generated by the study are available from the corresponding author on reasonable request].

**Code Availability Statement** This code/software will be made available on reasonable request. [Authors' comment: The code and software generated during the current study are available from the corresponding author on reasonable request].